\documentclass[journal=jctcce,manuscript=article]{achemso}
\pdfoutput=1
\usepackage{algorithmic}
\usepackage{simplewick}

\newlength\myindent
\def\ie{i.e., }

\def\bravert{\egroup\,\vrule\,\bgroup}

{\catcode`\|=\active
  \gdef\Twoint#1{\left(\mathcode`\|"8000\let|\bravert {#1}\right)}}
{\catcode`\|=\active
  \gdef\Braket#1{\left<\mathcode`\|"8000\let|\bravert {#1}\right>}}

\newcommand{\beq}{\begin{equation}}
\newcommand{\eeq}{\end{equation}}
\newcommand{\beqa}{\begin{eqnarray}}
\newcommand{\eeqa}{\end{eqnarray}}
\newcommand{\bea}{\begin{array}}
\newcommand{\eea}{\end{array}}

\newcommand{\bef}{\begin{figure}}
\newcommand{\ef}{\end{figure}}
\newcommand{\bc}{\begin{center}}
\newcommand{\ec}{\end{center}}
\newcommand{\bt}{\begin{table}}
\newcommand{\et}{\end{table}}
\newcommand{\btb}{\begin{tabular}}
\newcommand{\etb}{\end{tabular}}

\newcommand{\map}{{\emph{a priori }}}

\title {\sc{The Integral- and Intermediate-Screened Coupled-Cluster Method}}
\author{L.K.~S{\o}rensen}
\affiliation{Department of Chemistry - {\AA}ngstr{\"o}m Laboratory,
             Uppsala University,
             S-75105 Uppsala,
             Sweden}
\email{lasse.kraghsorensen@kemi.uu.se}


\begin{document}
%%%%%%%%%%%%%%%%%%%%%%%%%
\begin{abstract}
 We present the formulation and implementation of
the integral- and intermediate-screened coupled-cluster method (ISSCC).
The IISCC method gives a simple and rigorous
integral and intermediate screening (IIS) of the
coupled-cluster method and will
significantly reduces the scaling for all orders of the CC hierarchy
exactly like seen for the integral-screened configuration-interaction method (ISCI).
The rigorous IIS in the IISCC gives a robust and adjustable error control
which should allow for
the possibility of converging the energy without any loss of accuracy
while retaining low or linear scaling at the same time.
The derivation of the IISCC is performed
in a similar fashion as in the ISCI
where we show that the tensor contractions
for the nested commutators are separable
up to an overall sign and that this
separability can lead to a rigorous IIS.
In the nested commutators the integrals are
screened in the first tensor contraction and the intermediates are screened
in all successive tensor contractions.
The rigorous IIS will lead to linear scaling
for all nested commutators
for large systems in a similar way as seen for the ISCI.
The reduced scaling can in this way be obtained without any range dependent
parameter for the interaction, unlike other low
scaling methods and is therefore also suitable
for charge separated systems.
It is expected that the IISCC, just like the ISCI,
can use dramatically larger
orbital spaces combined with large CC expansions
as compared with traditional CC methods.
Due to the size extensive nature of the CC method
the IISCC should not only be applicable to
few electron systems but also to large molecular
system unlike the ISCI. The IISCC can here
either be used as a stand alone method or
combined with other linear scaling approaches
like the fragmentation method or other
methods which introduces a range dependent
interaction screening.
Three different ways to obtain IIS for the CC 
method is shown and the algorithm 
for two of these are shown and discussed. The first algorithm
is very similar to the regular CC method where
intermediates are collected and an IIS is used for
every tensor contraction. In the second method
all contractions in the nested commutators are performed in a single step.
Here an IIS is used before every multiplication 
with the amplitudes for a cluster operator which will
effectively remove the vast majority of all
multiplications with the amplitudes.

\end{abstract}

\maketitle
%\tableofcontents{}

\section{Introduction}
%%%%%%%%%%%%%%%%%%%%%%%%%%%%%%%%%%%%%%%%%%%%%%%%
\label{SEC:intro}

The coupled-cluster method (CC) originally introduced in
nuclear physics by Coester and K{\"u}mmel \cite{cc1,cc2}
and some years later picked up by {\v C}{\'i}{\v z}ek, Paldus and Shavitt \cite{cc3,cc4}
started in the late 70s to early 80s to replace the 
configuration-interaction method (CI) \cite{kellner_ci1,kellner_ci2,hyll_ci} as the workhorse
in accurate quantum chemistry calculations. The replacement was quite obvious
since the non-linear wave function ansatz of the 
CC method gives a significantly more
compact representation of the wave function in 
comparison to linear ansatz in CI. The success
of the CC method
in accurately capturing the correlation energy
is largely connected to the ability
to include important disconnected correlation effects
already at the CCSD level. The CC ansatz is therefore
still the most efficient wave function 
parameterization known for capturing the
dynamic electron correlation effects.
Here the CCSD(T) method \cite{ccsdpt} has become known as the
"golden standard" for closed shell systems near the 
equilibrium geometry in quantum chemistry
due to the methods ability to accurately capture
the triple excitations perturbatively and
quadruples via disconnected correlation effects.

Due to the steep scaling of wave function methods the size of the
systems for which the standard CC method is
applicable is limited. A great deal of work has therefore
been placed on a large scale parallelization of the CCSD and CCSD(T)
equations \cite{janowski_2007,janowski_2008,Harding_2008,nwchem,cc_par_2013,nwchem_2014}
which either seeks to reduce the memory requirement
or communication between nodes while still
trying to have an efficient code. While the
massive parallelization of the CC equations 
enables the ability of calculating slightly larger
systems in a reasonably time the fundamental
problem of the steep scaling cannot be overcome
simply by the use of more computational power
alone and a different strategy, which reduces the
scaling, for very large systems
is therefore needed.

During the last decade the interest in 
CC methods with reduced or linear
scaling \cite{werner_lin_sca,werner_lin_sca2,bartlett_lin,poul_lin_sca,poul_lin_sca2,neese_lin2,incremental,stoll,li_lin_sca_09,li_lin_sca_10,Kobayashi_lin_sca} 
has therefore grown significantly. These methods
all in some way relies on long
range screening in local orbitals as proposed by Pulay \cite{pulay_lin_sca,pulay_lin_sca2}.
The integral- and intermediate-screened coupled-cluster method (IISCC),
as proposed here, will also rely on local orbitals for an improved reduced
scaling but unlike all other methods will not
introduce domains in order to achieve a range dependent screening.
The IISCC method will instead follow the integral
screening (IS) of the recently proposed 
integral-screened configuration-interaction method (ISCI) \cite{isci}
which introduces a simple and rigorous IS by looping over the integral
indices first in the tensor contractions. 
The IS in the ISCI therefore also minimizes the
number of times an integral will have to be calculated which is essential in very 
large basis sets where the integrals no longer can be stored in memory
but have to be calculated on the fly. 
The energy in the IISCC, like the ISCI, 
can be more accurately approximated than other low scaling methods
since the screening is directly built into the method and
therefore does not need to introduce
domains for a distance dependent interaction screening
in order to obtain a low scaling. Unlike all
other low scaling methods where the scaling locally in their
domains are the same as in the regular CC method the IISCC
will still show reduced scaling within these domains and is therefore not as sensitive
to the number of basis functions included. 
Strong-field time-dependent calculations with the ISCI
has shown that the ISCI method is capable of using
very large basis sets even without any form of
parallelization \cite{Larsson}. It is therefore expected
that the IISCC method will show similar qualities as the ISCI method
since the IIS or IS, which is essential in the scaling reduction,
is the same for both methods.

The main reason why CC and 
M{\o}ller-Plesset (MP) perturbation theory \cite{mppt} is central
in the development of low scaling methods and not CI is due to 
the CC and MP methods being size-extensive \cite{size-exten1,size-exten2}.
The central problem in transferring the IS from the ISCI to the IISCC is the non-linear
nature of the CC equations which normally results in the construction 
of intermediates in the CC method. While the non-linear
ansatz in CC does not resemble the linear ansatz in CI the
difference between these from an implementation viewpoint are
not significantly different since both methods can be reduced
to a series of tensor contractions where the main difference
is the contractions and collection of intermediates that
occur in the CC method.

In the following the ISCI and the idea of IS behind this
is briefly repeated since this can immediately be used
in the IISCC to show that all contractions that are within
the projection manifold can be handled in the exactly
same manner as in the ISCI method. This will help to substantiate
that the results shown for the ISCI \cite{isci} is almost
directly transferable to the IISCC. Thereafter will the CC equations
briefly be presented along with the idea behind the IISCC method
where also the great similarities between the ISCI and IISCC methods
will be illustrated. In order to obtain not only IS but
integral and intermediate screening (IIS), where a
screening is performed every time an integral or intermediate is
multiplied with an amplitude,
a discussion about the best way of collecting intermediates
for large systems in large basis sets is undertaken.
This will be followed by a derivation of the
IISCC working equations where three very different approaches
to obtain a simple and rigorous IIS is discussed.
Here it will be shown that a complete IIS is possible 
for nested commutators in several very different ways 
and that the scaling reduction
seen in the ISCI also will be possible for the IISCC.
The rewriting of the Hamiltonian, like in the ISCI,
will also make the CC equations separable up to 
an overall sign and is what enables the IIS.

Despite the IISCC method not yet being implemented
a pseudo code of how an implementation can be performed
in two different ways
is shown in order to show how IIS and reduced scaling of
the IISCC method is possible without any need to introduce
domains for a range dependent screening. The first way
will be very similar to the usual way the CC equations
are solved by using intermediates 
but where every tensor contraction is performed
in a way similar to the ISCI. This will give an IIS
in all the usual terms in the CC equations and therefore
reduce the scaling of these individually. If the
rate limiting step is the loop over the integrals
all excitations below doubles will have a cost
similar to the ISCISD. If, however, triples or higher are
included the construction of the intermediates
will increase the cost of the limiting step. The second
way will perform all contractions in the
nested commutators in a single step and therefore
not explicitly construct any intermediates. While
this formally will give the wrong scaling the IIS
can be used to reduce this since a higher screening
threshold can be used for nested commutators. The 
second advantage of performing all contractions in 
a single step and not collecting intermediates is that
the rate limiting step will not be affected by including
higher excitations.

While the IISCC here is derived as a separate method
this can be combined with both the fragmentation method
and the local coupled-cluster methods
to give local screening and enable the use of much larger
orbital spaces in the correlation of the different domains.
Exactly like in the ISCI will it also be possible in the IISCC to
introduce a range dependent interaction screening in
a simple way to reduce the loop over the integrals.

\section{Theory}
\label{SEC:theo}
In order to derive the IISCC equations in a way where the
similarities to the ISCI equations is evident the 
CI and ISCI \cite{isci} methods will briefly be recapitulated.
The CC method will follow where the usual approach
to collecting intermediates will be revisited. Here
it will be shown that 
a combined IIS
in the nested commutators in 
the CC method can be realized in several different ways 
by using the separation
techniques from the ISCI. For very large systems
the size of the intermediates will become significant
and therefore impossible to
store in memory or on disc \cite{krcc}. Although the large
size of the intermediates can be overcome by a stepwise
construction of the intermediates alternatives 
where no intermediates are explicitly constructed
but which still have a simple and rigorous IIS
will be shown.

Throughout the indices p,q,r,s\ldots are general indices
running over both the occupied orbitals (O) and the virtual
orbitals (V) while a,b,c,d\ldots will be used for the virtual
orbitals and i,j,k,l\ldots for the occupied orbitals. 
Both the ISCI and IISCC formulation are based
on a single reference determinant.
Excitation and de-excitation
terms with respect to the reference determinant can be defined which
means that any creation operator $\hat a^{\dagger}$
with indices a,b,c\ldots is an excitation operator while the
indices i,j,k\ldots will give a de-excitation operator and
the opposite is true for the annihilation operator $\hat a$.
The de-excitation terms are found in both the 
Hamiltonian and the intermediates, if the latter are constructed,
and that the de-excitation indices will have
to be matched by excitation terms from the CC or CI excitation operator
to give non-zero contributions.

\subsection{Configuration Interaction (CI)}

In CI the wavefunction $|{\mathbf C} \rangle$ 
is constructed as a linear combination of Slater determinants which corresponds to a
parameterization where 
an excitation operator $\hat X$ works on a reference determinant $|0 \rangle$,
\beq
\label{ci_ansatz}
 | {\mathbf C} \rangle = \hat C | 0 \rangle = \sum_i c_i \hat X_i | 0 \rangle ,
\eeq
which generates all possible determinants.
The expansion coefficients $c_i$ are found
by a variational optimization of the expectation value
of the electronic energy which is equivalent
to an eigenvalue problem for the coefficients and energy
\beq
\label{hcec}
{\mathbf H} | {\mathbf C} \rangle = E | {\mathbf C} \rangle.
\eeq 
The FCI in Eq. \ref{ci_ansatz} is exact in a complete basis and the best
approximation in an incomplete basis but only possible
for systems with few electrons in modest basis sets 
due to an exponential scaling.
A hierarchy
\beq
\label{ci_hierarchy}
\hat C = \sum_{i=0}^N \hat C_i 
                     = c_0 + \sum_{a,i}^{V,O} c_{i}^{a} \hat{a}_a^\dagger \hat{a}_i 
                     + \sum_{a>b,i>j}^{V,O} c_{ij}^{ab} \hat{a}_a^\dagger \hat{a}_b^\dagger \hat{a}_i \hat{a}_j
                     + \ldots  ,
\eeq
where the excitation operator is divided into excitation operators
with particle rank (see Eq. \ref{particle_rank} in Appendix \ref{ex_class_form}) 
spanning from zero to $N$ is therefore introduce, where $N$ is the number of particles.

The eigenvalue problem in Eq. \ref{hcec} is solved
iteratively by repeatedly applying the Hamiltonian to an
approximate eigenvector ${\mathbf v}$ to give the linearly transformed
approximate eigenvector ${\mathbf \sigma}$
\beq
\label{sigma}
{\mathbf H} {\mathbf v} = {\mathbf \sigma},
\eeq
which is known as the $\sigma$-vector. 
An efficient solution to the $\sigma$-vector step
is central in CI. 
Apart from the $\sigma$-vector step an optimization step where
typically a Davidson \cite{davidson_ci} or Lanczos \cite{lanczos} algorithm is used to
find the step for the new approximate eigenvector ${\mathbf v}$.

\subsection{The ISCI and integral screening}
\label{screen}

The CI Hamiltonian is typically very sparse where the first
part comes from trivially zero matrix elements, the Slater-Condon rules,
and the second part from many very small elements
which can be considered numerically zero. The position of
the trivially zero matrix elements are easy to find but
the numerical zeros in the CI Hamiltonian are significantly harder to predict.
In the usual low or linearly scaling methods the numerical zeros
are found by transforming to local orbitals and introducing
a distance dependent screening. The distant dependent
screening can also be used in the IISCC but will not at the
present stage.
Another type of sparsity also exist, that is present also for large
matrix elements, where the matrix element is a sum of one or many integrals since these matrix elements will
contain many numerically zero integrals. Since these
small integrals takes time to calculate but makes no contribution
to any properties it is
desirable to screen away all the  
small integrals, even those inside large matrix
elements. The central idea in the ISCI is to minimize
the number of times an integral have to be calculated and 
completely screen away all small integrals in a simple way.

Any matrix element in the Hamiltonian ($H_{pq}$) can be written as a sum of integrals $I_r$
\beq
\label{h_int_sum}
H_{pq} = \sum_r I_r .
\eeq
Matrix multiplication is distributive so the $\sigma$-vector
step in Eq. \ref{sigma} can be written as
\beq
\label{sigma_int}
\sum_t^{I_{all}} {\mathbf H}_t {\mathbf v} = {\mathbf \sigma}, \qquad 
({\mathbf H}_t){_{pq}} = \left\{\begin{array}{l} 0 \\ I_t \end{array}, \right.
\eeq
where the sum is over all integrals $I_{all}$. Each matrix element in
${\mathbf H}_t$ can now only take the values $0$ or $I_t$ and ${\mathbf H}_t$ is
therefore extremely sparse.
By introducing a predefined threshold parameter $\epsilon$ for the 
IS the
summation in Eq. \ref{sigma_int} can be split into two sums
\beq
\label{sigma_split}
\sum_t^{I_{all}} {\mathbf H}_t = \sum_l^{I_{large}} {\mathbf H}_l + \sum_s^{I_{small}} {\mathbf H}_s, 
\qquad |I_s | < \epsilon \le | I_l |.
\eeq
From Eq. \ref{sigma_split} introducing an IS which affects all
matrix elements containing a given integral $({\mathbf H}_t)$ is very simple
since this only requires knowing the value of $I_t$.
The integral $I_t$ in principle only needs to be
calculated once in order to screen away $I_t$ in all of
${\mathbf H}$. In this way the number of times
an integral have to be calculated is minimized and a complete IS,
even of large matrix elements, is easily accomplished.
This type of IS 
is ideal for very large basis sets where the integrals 
cannot be stored in memory but have to be calculated on the fly. 

If $I_{small} >> I_{large}$
and if there is a fast way of finding and multiplying the
non-zero elements in ${\mathbf H}_t$ with the elements
in ${\mathbf v}$ then solving the CI problem using
Eq. \ref{sigma_split} can give a significant reduction in
the scaling \cite{isci}. Using local orbitals $I_{small}$
will grow as $N^4$, in a Gaussian basis set, with system size while $I_{large}$
only will grow as $N$ if the system is sufficiently spatially
extended since all integrals between orbitals
sufficiently far apart will be below $\epsilon$. 
The reduction in scaling will depend on the desired
accuracy hence $\epsilon$. The IS in the ISCI is unlike
all the IS in all other methods which rely
on a distance dependent screening
for spatially extended systems in local orbitals 
\cite{werner_lin_sca,werner_lin_sca2,bartlett_lin,poul_lin_sca,poul_lin_sca2,neese_lin2,incremental,stoll,li_lin_sca_09,li_lin_sca_10,Kobayashi_lin_sca,lin_sca_mrci,pulay_lin_sca,pulay_lin_sca2}.

Due to the focus on having a simple and rigorous IS the ISCI
does not follow the usual strategy for CI algorithms 
where matrix elements are constructed from a sum of integrals.
The ISCI is instead completely driven by the integrals,
which will have to be in the outer loop in the $\sigma$-vector step,
and no matrix elements will ever be constructed. 

\subsubsection{Hamiltonian}

A time-independent non-relativistic one- and two-body Hamiltonian,
where the one particle functions can be related by the spin-flip
operator, is here used but the ISCI formalism is completely general
so any other Hamiltonian could easily be inserted instead.
In second quantization the Hamiltonian is a sum
of one- and two-body operators,
\beqa
\label{index_re}
\hat H &=& \sum_{pq} h_{pq} (\hat{a}_{p{\alpha}}^\dagger \hat{a}_{q{\alpha}} + \hat{a}_{p{\beta}}^\dagger \hat{a}_{q{\beta}}) \\
       &+& \sum_{p>r,q>s} (g_{psrq} - g_{pqrs})( \hat{a}_{p{\alpha}}^\dagger \hat{a}_{r{\alpha}}^\dagger \hat{a}_{q{\alpha}} \hat{a}_{s{\alpha}}
        +                                   \hat{a}_{p{\beta}}^\dagger \hat{a}_{r{\beta}}^\dagger \hat{a}_{q{\beta}} \hat{a}_{s{\beta}}) \nonumber \\
       &-& \sum_{pqrs} g_{pqrs} \hat{a}_{p{\alpha}}^\dagger \hat{a}_{r{\beta}}^\dagger \hat{a}_{q{\alpha}} \hat{a}_{s{\beta}}. \nonumber
\eeqa
where $h_{pq} = \langle \phi_p | h | \phi_q \rangle$ and
$g_{pqrs} = \langle \phi_p \phi_r | g | \phi_s \phi_q \rangle $
are the integrals associated with the one- and two-body operators.
The Hamiltonian in Eq. \ref{index_re}, 
is normal- and spin-ordered and index-restricted
unlike the UGA Hamiltonian \cite{paldus_uga,shavitt_uga}.
The order chosen for the spin order and index restriction are
arbitrary and any other order would not change the
ISCI method since 
any sign change in the integrals in the Hamiltonian
would be compensated by an overall sign change
in Eq. \ref{final}. 

By choosing a canonical orbital ordering, 
i.e., occupied before virtual orbitals, 
it can be shown \cite{isci}
that any term in the Hamiltonian in Eq. \ref{index_re} 
can be written as
\beq
\label{exdx}
\hat H_{any} = \hat C_{\alpha}^{ex} \hat C_{\alpha}^{dx} \hat C_{\beta}^{ex} \hat C_{\beta}^{dx} 
               \hat A_{\alpha}^{dx} \hat A_{\alpha}^{ex} \hat A_{\beta}^{dx} \hat A_{\beta}^{ex}.
\eeq
In Eq. \ref{exdx} the operator is written in 
terms of strings of second quantized operators with indices ordered according
to the order in the Hamiltonian in Eq. \ref{index_re}.
Here $\hat C$ and $\hat A$ are strings of 
creation and annihilation operators, $\alpha$ and $\beta$ the
spin and the $ex$ and $dx$ superscripts denotes if an operator is an
excitation or a de-excitation operator, respectively. 
$\hat C_{\alpha}^{ex}$ is therefore the excitation part of the 
creation operators with $\alpha$ spin.  
The integrals have been omitted since these will be defined
from the strings in a given operator.

\subsubsection{The $\sigma$-vector step}

Any part in the $\sigma$-vector step in 
Eq. \ref{sigma} can be written like
\beq
\label{sig}
\hat \sigma_{c \alpha} \hat \sigma_{c \beta} \hat \sigma_{a \alpha} \hat \sigma_{a \beta} =
\hat C_{\alpha}^{ex} \hat C_{\alpha}^{dx} \hat C_{\beta}^{ex} \hat C_{\beta}^{dx}
\hat A_{\alpha}^{dx} \hat A_{\alpha}^{ex} \hat A_{\beta}^{dx} \hat A_{\beta}^{ex}
\hat v_{c \alpha} \hat v_{c \beta} \hat v_{a \alpha} \hat v_{a \beta}.
\eeq
Rearranging Eq. \ref{sig} using the elementary anti-commutation rules
for second quantized operators 
\beq
\label{final}
\hat \sigma_{c \alpha} \hat \sigma_{c \beta} \hat \sigma_{a \alpha} \hat \sigma_{a \beta} =
\hat C_{\alpha}^{ex} \hat A_{\alpha}^{dx} \hat v_{c \alpha}
\hat C_{\beta}^{ex} \hat A_{\beta}^{dx} \hat v_{c \beta}
\hat C_{\alpha}^{dx} \hat A_{\alpha}^{ex} \hat v_{a \alpha}
\hat C_{\beta}^{dx} \hat A_{\beta}^{ex} \hat v_{a \beta} (-1)^M ,
\eeq
where only transpositions 
which can give a sign change have been performed.
$M$ is then the number of transpositions needed to rearrange
the operators from Eq. \ref{sig} to Eq. \ref{final}.
It is seen that Eq. \ref{final} can be split into four parts: 
\beqa
\label{four}
\hat \sigma_{c \alpha} = \hat C_{\alpha}^{ex} \hat A_{\alpha}^{dx} \hat v_{c \alpha}, \\
\label{fourb}
\hat \sigma_{c \beta} = \hat C_{\beta}^{ex} \hat A_{\beta}^{dx} \hat v_{c \beta}, \\
\label{lasta}
\hat \sigma_{a \alpha} = \hat C_{\alpha}^{dx} \hat A_{\alpha}^{ex} \hat v_{a \alpha}, \\
\label{last}
\hat \sigma_{a \beta} = \hat C_{\beta}^{dx} \hat A_{\beta}^{ex} \hat v_{a \beta},
\eeqa
which each have to be fulfilled for a non-zero
contribution in the $\sigma$-vector calculation. 

Equations \ref{four}-\ref{last} have been solved \cite{isci} by
applying a Hamiltonian term 
to $\hat v$, which in the operator form
is identical to the excitation operator $\hat X$. Using
elementary operations for second quantized operators, the 
different parts of the $\sigma$-vector can be found.
For every non-zero operation the indices of the Hamiltonian, 
$\hat v$ and $ \hat \sigma$ is tabulated and used
in the $\sigma$-vector step since each of these indices 
will give a part of the integral
multiplied with the coefficients in $\hat v$ to $\hat \sigma$.

By combining the solutions from Eqs. \ref{four}-\ref{last}
the desired loop structure from Sec. \ref{screen} in the
$\sigma$-vector step can be achieved. By looping over the indices
for the integrals a simple and rigorous IS can be
accomplished by a simple if-statement which can 
be seen from the pseudo code where all integral below
$\epsilon $ are screened away in line \ref{gci3}.
\begin{algorithmic}[1]
\LOOP[ Integral indices ]
 \STATE{Fetch or calculate integral $I$}
 \IF{ $|I| > \epsilon $ } \label{gci3}
 \LOOP[ Matrix elements ]
  \STATE{Multiply integral with element in $\hat v$ to element in $\hat \sigma$}
 \ENDLOOP[ Matrix elements ]
 \ENDIF                                                                                                                                               
\ENDLOOP[ Integral indices ]                                                                                                                          
\end{algorithmic}                                                                                                                                     
The integral loop constitutes the minimum in the number
of parameters that have to be looped over in 
any calculation if no assumption about a long range
decay of the integrals is invoked.
The IS will give a gradual scaling reduction, until linear, 
in the inner loop where the integral is multiplied with a CI coefficient for 
spatially extended systems expressed in local orbitals \cite{isci}.
The aim is to rewrite the CC equations in a similar way where
the outer loop consist over the integral indices even
for nested commutators since this also will give a simple
and rigorous IS in CC.

\subsection{The Coupled-Cluster Approach}
\label{cc}

We will now turn to the single-reference CC theory
in order to show the different wave function ansatz in comparison
to CI and to demonstrate the reason why intermediates are
collected in the CC method. 
In CC theory the
wave function is parameterized by the exponential of an excitation operator
working on an $N$-particle reference function $| 0 \rangle$
\beq
|CC \rangle = \exp(\hat T ) | 0 \rangle .
\eeq
In the similarity-transformed formulation, the CC energy and amplitude
equations become
\beq
\label{ccenul}
\langle 0 | \exp( - \hat T ) \hat H \exp(\hat T ) | 0 \rangle = E
\eeq
and
\beq                                                                                                                                                  
\label{ccamp}                                                                                                                                         
\langle \mu | \exp( - \hat T ) \hat H \exp(\hat T ) | 0 \rangle = 0,                                                                                 
\eeq                                                                                                                                                  
respectively. It is seen that the  Baker-Cambell-Hausdorff expansion
%
%\begin{align}
\beq
\label{vcc_as_comm}
\Omega_{\hat T} = %\nonumber \\
 \langle 0 |\hat \tau^\dagger_{\mu}
(\hat H + [\hat H, \hat T]+\frac{1}{2!}[[\hat H,\hat T],\hat T]
+\frac{1}{3!}[[[\hat H,\hat T],\hat T],\hat T]
+\frac{1}{4!}[[[[\hat H,\hat T],\hat T],\hat T],\hat T])
| 0 \rangle 
\eeq
%\end{align}
%
terminates after the fourth order for the similarity-transformed 
Hamiltonian since the Hamiltonian is a two-particle operator
and $\hat T$ a pure excitation operator.

Including all excitations into $\hat T$  leads to the                                                                                         
full coupled-cluster model (FCC) which solves the Schr{\"o}dinger or Dirac                                                                            
equation in the defined one-electron basis. However, the                                                                                              
operator $\hat T$ is typically restricted like seen in
CI in Eq. \ref{ci_hierarchy} 
\beq                                                                                                                                                  
\label{texpansion}                                                                                                                                    
\hat T = \sum_{i=1}^m \hat T_i                                                                                                                        
                     = \sum_{a,i}^{V,O} t_{i}^{a} \hat{a}_a^\dagger \hat{a}_i 
                     + \sum_{a>b,i>j}^{V,O} t_{ij}^{ab} \hat{a}_a^\dagger \hat{a}_b^\dagger \hat{a}_i \hat{a}_j
                     + \ldots 
\eeq
where a truncation at $\hat T_2$ or $\hat T_3$ will give the familiar
CCSD or CCSDT models, respectively.

\subsubsection{Contractions and collection of intermediates}
\label{col_int}

In CC the focus is not only on capturing the direct
interaction, like in CI, but also the indirect interaction where the latter
appears as disconnected clusters or nested commutators in the BCH expansion in Eq. \ref{vcc_as_comm}.
Here in particular does the doubly nested commutator  
$\frac{1}{2!}[[\hat H,\hat T_2],\hat T_2]$
play the essential role in the numerical success of CC.
In the calculation of the nested commutators the usual procedure is
to construct intermediates. The two main reasons for
the construction of intermediates is to
obtain the correct scaling by not having to perform
multiple contractions between multiple tensors in
a single step and to reduce the prefactor
in the contractions by noting that there will be
several tensors with the same rank that has to be
contracted with the cluster operator and these
tensors can be added prior to the contraction
since the tensor product is distributive.
This was realized very early on but finding
the optimal way of contracting the tensors
even for CCSD was not easy \cite{scuseria_1986,scuseria_1988}.
The optimal way is, however, very dependent on the
size of the basis set and should in principle be optimized
separately for every calculation with for example 
a genetic algorithm\cite{hanrath_genetic}. The discussion about the
scaling and collection of intermediates is
not new and the wast majority of CC codes
will collect the intermediates in some way
for the reasons given here. The discussion is,
however, relevant to repeat once the IS in the
IISCC is introduced.

The contraction
processes between two operators will be written as
\beq
\label{cont_pro}
\hat O^{P,H} \otimes \hat T (XY) \Rightarrow \hat O^{P',H'}
\eeq
where $P$ is the number of particle indices still to be contracted in
operator $\hat O$, before this can be added to the
projection manifold $\Omega_{\hat T}$, $X$ is the number of particle indices contracted
and $P'= P-X$ is the remaining particle indices to be contracted and
likewise for the holes $H' = H- Y$. In Eq. \ref{cont_pro} it
is assumed that the contraction is always done with $\hat T$
and all contractions between the $N \Delta M$ classes, as defined in Appendix \ref{ex_class_form}, are performed.
Eq. \ref{cont_pro} should therefore read as the tensor
contraction between all operators in $\hat O^{P,H}$ and $\hat T(XY)$ added to
$\hat O^{P',H'}$ where $X$ and $Y$ defines the indices                                                                                                
contracted.                                                                                                                                           
In this way it is easy to address which contraction is taking place                                                                                   
and to have different contractions pointing to the same intermediate                                                                                  
with given $P,H, N,\Delta M_s,M_{\alpha \beta}$ indices or to see how contractions from different                                                               
nested commutators can be combined to a single intermediate.

From scaling considerations, collection and size of intermediates 
a priority for the order the contractions can be made
or optimized on an individual basis
with a genetic algorithm\cite{hanrath_genetic}.
Here the following contraction order is chosen
\beq
\label{conord}
P,H:2,2;2,1;1,2;2,0;1,1;1,0;0,2;0,1.
\eeq
where the contraction performed first will be the leftmost
possible contraction. This kind of contraction order
can be introduced since the order in which the contractions
in the nested commutators are performed only matter from
a scaling perspective and hence all permutations of
contraction orders can therefore be collected in one order. 
Hence for the doubly nested commutator for $\hat H^{1,1}$ 
the particle index is contracted before the hole index since
$1,0$ is before $0,1$ in the contraction order in Eq. \ref{conord}.
We have chosen to contract as many indices as possible and contract
particles before holes except for the $1,0$ and $0,2$ contractions.
For the $1,0$ and $0,2$ contraction we have chosen the large
basis set limit where $ V > O^2$. The choice for the large basis set
limit was not only chosen, since the long term aim is to use very
large basis sets, but was also motivated by the way the intermediates was collected
since the largest of the intermediates $\hat M^{2,0}$ will not have to be constructed.
An example of this comes from the important
double commutator $\frac{1}{2!}[[\hat H,\hat T_2],\hat T_2]$ where only
\beq
\label{h22_r}
\hat H^{2,2} \otimes \hat T (20) \Rightarrow \hat M^{0,2} \otimes \hat T (02) \Rightarrow \Omega_{\hat T}
\eeq
and not
\beq
\label{h22_w}
\hat H^{2,2} \otimes \hat T (02) \Rightarrow \hat M^{2,0} \otimes \hat T (20) \Rightarrow \Omega_{\hat T}
\eeq
will be calculated since the result of the contractions are the same
and Eq. \ref{h22_r} will give a more favorable scaling of the contractions $(V^2 O^4)$
in comparison to Eq. \ref{h22_w} $(V^4 O^2)$ and the intermediate $\hat M^{0,2}$
is significantly smaller than $\hat M^{2,0}$ even for reasonably sized basis sets.

For the nested commutator in Eq. \ref{h22_r} the scaling of the
contractions are $(V^2 O^4)$ if the intermediate $\hat M^{0,2}$ 
is constructed. If, however, the problem becomes so large
that the intermediate no longer can be stored this will
either have to be constructed piecewise or the 
contractions will have to be performed in a single step as
\beq
\label{single_step}
\hat H^{2,2} \otimes \hat T (20) \otimes \hat T (02) \Rightarrow \Omega_{\hat T}.
\eeq
Performing the contractions in a single step
as shown in Eq. \ref{single_step} will give
a scaling of $O^4 V^4$ and therefore seem
less favorable. When the contractions are
performed in a single step the contraction order
will, however, not matter so the opposite order
\beq
\label{single_step2}
\hat H^{2,2} \otimes \hat T (02) \otimes \hat T (20) \Rightarrow \Omega_{\hat T}.
\eeq
will give exactly the same scaling as in Eq. \ref{single_step}.

The second reason for the collection of intermediates is
to reduce the prefactor of the contractions by
not having repeated contractions of same rank tensors.
Continuing the example of contractions with $\hat H^{2,2}$
from Eq. \ref{h22_r} it is seen that $\hat H^{0,2}$ can be
added to the intermediate so the second contraction in Eq. \ref{h22_r}
and the direct $\hat H^{0,2}$ contraction
\beq
\hat H^{0,2} \otimes \hat T (02) \Rightarrow \Omega_{\hat T}
\eeq
can be combined to 
\beq
\label{h22_c}
\hat H^{2,2} \otimes \hat T (20) \Rightarrow (\hat M^{0,2} \oplus \hat H^{0,2}) \otimes \hat T (02) \Rightarrow \Omega_{\hat T}.
\eeq
Eq. \ref{h22_c} simply combines the two contractions into one
and thereby halving the prefactor since the same
type of contractions are not repeated. With the contraction
order in Eq. \ref{conord} a contraction from $\hat H^{1,2}$
would also be added to the $\hat M^{0,2}$ intermediate
\beq
\label{h12_i}
\hat H^{1,2} \otimes \hat T (10) \Rightarrow \hat M^{0,2}.
\eeq
In this way will a given tensor contraction only be performed once.

\subsection{The IISCC}

The aim in the formulation of the IISCC is to obtain a similar
IS to that of the ISCI in Sec. \ref{screen}. Once 
the IS has been introduced the problem of collecting
intermediates will be revisited where, depending on
the correlation level, it will be shown that it may be 
advantageous to divert from
the two reasons for the collection
of intermediates discussed in Sec. \ref{col_int}
since an IIS with the same accuracy but higher
screening threshold for the nested commutators
can be obtained if intermediates are not explicitly
collected. 
After discussing the 
the different strategies for screening and contractions
the de-excitation terms will be separated in the 
Hamiltonian and a set of separable CC equations similar
to those for CI in Eqs. \ref{four}-\ref{last} presented. 

By rewriting the nested commutator expression in Eq. \ref{vcc_as_comm}
in terms of tensor contractions
\beq
\label{tensor_cc}
\sum_{n=0}^4 \frac{1}{n!} \hat H \otimes \hat T^{\otimes n} \Rightarrow \Omega_{\hat T}
\eeq
where the similarity to CI, as seen in Eq. \ref{sigma},
for $n$ up to one is seen. The IS in CC will therefore
be introduced in the same way as in the ISCI in Eqs. \ref{h_int_sum}-\ref{sigma_split}
for the Hamiltonian. By introducing the IS in the tensor contractions
of Eq. \ref{tensor_cc} 
\beq
\label{cc_is}
\sum_t^{I_all} \sum_{n=0}^4 \frac{1}{n!} \hat H_t \otimes \hat T^{\otimes n} \Rightarrow \Omega_{\hat T}
\eeq
an expression, where there is a loop over
the integral, for which the integrals can 
be screened is obtained exactly like in Eq. \ref{sigma_split}. The problem with
the nested commutators or multiple tensor contractions
is that the single integral from the Hamiltonian
can be used in many contractions from the initial
contraction with the cluster operator to an intermediate
which means that the intermediate can have many
different non-zero contributions and that the
number of these non-zero contributions will
grow with every contraction. 

\subsubsection{Contractions and collection of intermediates in the IISCC}
\label{col_int2}

The problem with the growth in the number
of terms with every contraction for integrals
larger than $\epsilon$ in principle suggests
at least three different ways to perform the contraction
where also an intermediate screening can be obtained.
The first way would be similar to that   
normally used in CC theory where the intermediates are
explicitly constructed or at least piecewise constructed
and this would lead not only to an IS but also
an IIS.
In the second a partial intermediates would be constructed
from the elements of the previous contraction
where in each step the intermediate is screened
in the contractions. In the third way all
contractions are performed in a single step
as shown in Eq. \ref{single_step}
and both integrals and intermediates are screened. 
Pseudo algorithms for the
separable CC equation and with IIS will be
shown in Sec. \ref{SEC:imp}.

\paragraph{Collecting all intermediates:} 

The main problem when the intermediates are explicitly constructed or at least
piecewise constructed is the fact that these will
have to be stored in some way. The equations
will, however, be very similar to the usual CC methods.
The intermediates can then be constructed as shown in Eqs. \ref{h22_c}-\ref{h12_i}
where in the contraction of the intermediate $\hat M^{0,2}$
with the cluster operator also can be screened in
the exact same manner as $\hat H^{0,2}$ can.
The main problems with adding the intermediates
in the traditional way, even if this is piecewise,
is that the intermediates will have to be stored
and several contractions will point to
the same intermediate and the structure of the equations
will therefore be more complicated. While at the
doubles level none of the intermediates will
be larger than the integral block and the
looping over the intermediates will therefore
not increase the outer loop in the contractions. However,
once triples or higher is included not only
will the size of the intermediates increase but
the outer loop in the contractions will also
increase in scaling and since this should be the
limiting step in the IISCC this will be a problem.
 
\paragraph{Collecting intermediates from single integral:} 

In the second way there will be an IS in the outer loop
where once an integral is above the IS threshold $\epsilon$
it will be multiplied with the CC amplitudes to an
intermediate
\beq
\label{h22_ex}
\hat H^{2,2}_{t} \otimes \hat T (20) \Rightarrow \hat M^{0,2}_{t} \qquad |I_t| > \epsilon.
\eeq
The intermediate $\hat M^{0,2}_{t}$ will contain a lot fewer elements
than the whole intermediate $\hat M^{0,2}$ and can therefore
significantly easier be looped over
\beq
\label{h22_ex2}
\hat M^{0,2}_{t} \otimes \hat T (02) \Rightarrow \Omega_{\hat T}
\eeq
and screened. In the double commutator the storage of the
intermediate will not be a problem but while formally only
the prefactor will increase, since the intermediates
is not collected from different contractions and integrals,
the number of multiplication will be close to when
all contractions are performed in a single step as
shown in Eq. \ref{single_step} and below.
If the number of small integrals
and amplitudes is large in comparison to the large ones
then the prefactor increase will not
matter since the IIS will reduce this significantly.
Furthermore for higher nested commutators a 
larger $\epsilon$ can be used in the first contractions without compromising
the accuracy since these will be multiplied with
multiple amplitudes. Storage problems can, however,
still occur for higher nested commutators. The order
of contractions will here affect both the size
of the intermediates and the scaling in the usual
manner as described in Sec. \ref{col_int}.

\paragraph{All contractions in a single step:} 

In the third way all contractions are performed
simultaneously
\beq
\label{single_step_t}
\hat H^{2,2}_t \otimes \hat T (20) \otimes \hat T (02) \Rightarrow \Omega_{\hat T}.
\eeq
The main difference of this approach in comparison 
to the second approach is the fact that there is no
need for storing the intermediates and the order
of contraction does not matter with respect to the scaling.
The simultaneous performance of contractions
is usually not carried out since this complicates the algorithm
and gives an incorrect scaling since terms are
not collected after each contraction as shown in Eq. \ref{single_step}. The advantages
are that the contraction order in principle does not matter,
the efficiency of the loop structure can be
optimized if some knowledge of the size and dimension
of the different parts of $\hat T$ is used,
a higher screening threshold for nested commutators
can be used without any loss of accuracy
and an IIS will ensure that only very few
contributions from the nested commutators
will be calculated.

\subsubsection{The IISCC Hamiltonian and intermediates}

Unlike in the CI not all de-excitation indices are contracted
in a single step in CC. For the IISCC we will therefore
assume that the Hamiltonian will at most contain two
particle operators which means that the down rank 
will at most be two. In higher order CC the particle rank
of the intermediates can be larger than two but the
down rank will, however, always be lower than two
since any contraction between the Hamiltonian and
the cluster operator will reduce the down rank.
A general mixed operator $\hat O$ which covers both
the Hamiltonian and the intermediates can be written as
\beq
\label{iscc_ham}
\hat O_{any} = \hat C_{\alpha}^{ex} \hat C_{\alpha}^{dx1} \hat C_{\alpha}^{dx2} \hat C_{\beta}^{ex} \hat C_{\beta}^{dx1} \hat C_{\beta}^{dx2}
               \hat A_{\alpha}^{dx1} \hat A_{\alpha}^{dx2} \hat A_{\alpha}^{ex} \hat A_{\beta}^{dx1} \hat A_{\beta}^{dx2} \hat A_{\beta}^{ex}
\eeq
where any integral or intermediate will be defined
from the operator strings. In the general operator
expression in Eq. \ref{iscc_ham} the particle
rank can be any order but the down rank can
maximally be two since the de-excitation terms
are now explicitly written as indicated by the 1 and 2
in the $dx$ superscript. $\hat O_{any}$ can therefore
be used for any two-particle Hamiltonian and intermediate.
For the Hamiltonian part the index restricted and normal- and
spin-ordered Hamiltonian, exactly like
in the ISCI in Eq. \ref{index_re}, is used and the
intermediates are arranged likewise. Any term in the
mixed operator in Eq. \ref{iscc_ham} will therefore
consist of a sum of the integrals from the Hamiltonian
and the coefficient from the intermediate.

\subsubsection{Rigorous integral and intermediate screening of nested commutators}
\label{riis}

By combining the general tensor contractions from the
nested commutators in Eq. \ref{tensor_cc} with the 
general operator from Eq. \ref{iscc_ham} a general
contraction scheme similar to that in CI in Eq. \ref{sig}
can be written for any contraction in the nested commutators
\beq
\label{sigcc}
\hat C_{\alpha}^{ex} \hat C_{\alpha}^{dx1} \hat C_{\alpha}^{dx2} \hat C_{\beta}^{ex} \hat C_{\beta}^{dx1} \hat C_{\beta}^{dx2}
\hat A_{\alpha}^{dx1} \hat A_{\alpha}^{dx2} \hat A_{\alpha}^{ex} \hat A_{\beta}^{dx1} \hat A_{\beta}^{dx2} \hat A_{\beta}^{ex} 
(\hat t_{c \alpha} \hat t_{c \beta} \hat t_{a \alpha} \hat t_{a \beta})^{n} =
\hat \Omega_{c \alpha} \hat \Omega_{c \beta} \hat \Omega_{a \alpha} \hat \Omega_{a \beta}.
\eeq
Exactly like in CI can Eq. \ref{sigcc} be reordered
\beq
\label{sigcc_reo}
\hat C_{\alpha}^{ex} \hat A_{\alpha}^{dx1} \hat A_{\alpha}^{dx2} \hat t_{c \alpha}^n
\hat C_{\beta}^{ex} A_{\beta}^{dx1} \hat A_{\beta}^{dx2} \hat t_{c \beta}^n
\hat C_{\alpha}^{dx1} \hat C_{\alpha}^{dx2} \hat A_{\alpha}^{ex} \hat t_{a \alpha}^n
\hat C_{\beta}^{dx1} \hat C_{\beta}^{dx2} \hat A_{\beta}^{ex} t_{a \beta}^n (-1)^M =
\hat \Omega_{c \alpha} \hat \Omega_{c \beta} \hat \Omega_{a \alpha} \hat \Omega_{a \beta},
\eeq
for any nested commutator. Just like in CI in Eq. \ref{sigcc_reo}
are the CC equations separable up to an overall sign exactly
like in Eqs. \ref{four}-\ref{last}
\beqa
\label{four_cc}
\hat C_{\alpha}^{ex} \hat A_{\alpha}^{dx1} \hat A_{\alpha}^{dx2} \hat t_{c \alpha}^n = \hat \Omega_{c \alpha}, \\
\label{four_cc2}
\hat C_{\beta}^{ex} A_{\beta}^{dx1} \hat A_{\beta}^{dx2} \hat t_{c \beta}^n = \hat \Omega_{c \beta}, \\
\label{four_cc3}
\hat C_{\alpha}^{dx1} \hat C_{\alpha}^{dx2} \hat A_{\alpha}^{ex} \hat t_{a \alpha}^n = \hat \Omega_{a \alpha}, \\
\label{final_cc}
\hat C_{\beta}^{dx1} \hat C_{\beta}^{dx2} \hat A_{\beta}^{ex} t_{a \beta}^n = \hat \Omega_{a \beta}.
\eeqa
For $n$ equal to zero or one the
contractions between the Hamiltonian and the cluster operator
can be performed in the exact same way as in CI while for
two, three and fourfold nested commutators special care
have to be taken.

Since the Hamiltonian is a number conserving two-particle operator
it is sufficient to show how the contractions can be performed
for a doubly nested commutator where two indices of an operator 
with same spin and type is contracted after each other. 
For a doubly nested commutator for $\hat H_{202}^{0,2}$ where
the following contraction are identical this can from Eq. \ref{four_cc3} be written as
\beq
\label{2foldex}
\hat C_{\alpha}^{dx1} \hat C_{\alpha}^{dx2} \hat A_{\alpha}^{ex} t_{a \alpha}^2 =
\hat C_{\alpha}^{dx1} \hat C_{\alpha}^{dx2} \hat A_{\alpha}^{ex} t_{a \alpha}^{(1)} t_{a \alpha}^{(2)} =
\hat A_{\alpha}^{ex} \hat C_{\alpha}^{dx1} \hat C_{\alpha}^{dx2} t_{a \alpha}^{(1)} t_{a \alpha}^{(2)} (-1)^M 
\eeq
since for doubly nested commutators where the two indices are contracted separately,
unlike in CI, there can be no internal contractions.
The number $x$ in $t_{a \alpha}^{(x)}$ only shows the order
of appearance in the nested commutators. 
The contractions with the cluster operator can then be resolved as
\beq
\label{diff_con}
\contraction{\hat A_{\alpha}^{ex} \hat C_{\alpha}^{dx1} \hat C_{\alpha}^{dx2} \hat t_{a\alpha}^{(1)} \hat t_{a\alpha}^{(2)} =
\hat A_{\alpha}^{ex} ( }{\hat C}{_{\alpha}^{dx1}\hat C_{\alpha}^{dx2}}{\hat t}
\contraction[2ex]{\hat A_{\alpha}^{ex} \hat C_{\alpha}^{dx1} \hat C_{\alpha}^{dx2} \hat t_{a\alpha}^{(1)} \hat t_{a\alpha}^{(2)} =
\hat A_{\alpha}^{ex} ( \hat C_{\alpha}^{dx1}}{ \hat C}{_{\alpha}^{dx2} \hat t_{a\alpha}^{(1)} }{\hat t}
\contraction{\hat A_{\alpha}^{ex} \hat C_{\alpha}^{dx1} \hat C_{\alpha}^{dx2} \hat t_{a\alpha}^{(1)} \hat t_{a\alpha}^{2)} =
\hat A_{\alpha}^{ex} ( \hat C_{\alpha}^{dx1} \hat C_{\alpha}^{dx2} \hat t_{a\alpha}^{(1)} \hat t_{a\alpha}^{(2)}+
                       }{\hat C}{_{\alpha}^{dx1} \hat C_{\alpha}^{dx2} \hat t_{a\alpha}^{(1)} }{\hat t}
\contraction[2ex]{\hat A_{\alpha}^{ex} \hat C_{\alpha}^{dx1} \hat C_{\alpha}^{dx2} \hat t_{a\alpha}^{(1)} \hat t_{a\alpha}^{(2)} =
\hat A_{\alpha}^{ex} ( \hat C_{\alpha}^{dx1} \hat C_{\alpha}^{dx2} \hat t_{a\alpha}^{(1)} \hat t_{a\alpha}^{(2)}+
                       \hat C_{\alpha}^{dx1} }{\hat C}{_{\alpha}^{dx2} }{\hat t}
\hat A_{\alpha}^{ex} \hat C_{\alpha}^{dx1} \hat C_{\alpha}^{dx2} \hat t_{a\alpha}^{(1)} \hat t_{a\alpha}^{(2)} =
\hat A_{\alpha}^{ex} ( \hat C_{\alpha}^{dx1} \hat C_{\alpha}^{dx2} \hat t_{a\alpha}^{(1)} \hat t_{a\alpha}^{(2)}+
                       \hat C_{\alpha}^{dx1} \hat C_{\alpha}^{dx2} \hat t_{a\alpha}^{(1)} \hat t_{a\alpha}^{(2)})
\eeq
where the two contractions gives the same result
\beq
\label{sum_con}
\contraction{\hat A_{\alpha}^{ex} \hat C_{\alpha}^{dx1} \hat C_{\alpha}^{dx2} \hat t_{a\alpha}^{(1)} \hat t_{a\alpha}^{2)} =
2 \hat A_{\alpha}^{ex}
                       }{\hat C}{_{\alpha}^{dx1} \hat C_{\alpha}^{dx2} \hat t_{a\alpha}^{(1)} }{\hat t}
\contraction[2ex]{\hat A_{\alpha}^{ex} \hat C_{\alpha}^{dx1} \hat C_{\alpha}^{dx2} \hat t_{a\alpha}^{(1)} \hat t_{a\alpha}^{(2)} =
2 \hat A_{\alpha}^{ex}
                       \hat C_{\alpha}^{dx1} }{\hat C}{_{\alpha}^{dx2} }{\hat t}
\hat A_{\alpha}^{ex} \hat C_{\alpha}^{dx1} \hat C_{\alpha}^{dx2} \hat t_{a\alpha}^{(1)} \hat t_{a\alpha}^{(2)} =
2 \hat A_{\alpha}^{ex}
                       \hat C_{\alpha}^{dx1} \hat C_{\alpha}^{dx2} \hat t_{a\alpha}^{(1)} \hat t_{a\alpha}^{(2)}.
\eeq
The doubly nested contraction shown in Eq. \ref{sum_con} will appear
for 
\beq
\label{H02_double}
\frac{1}{2!} [[ \hat H_{202}^{0,2} , \hat T], \hat T] \Rightarrow \Omega_{\hat T}
\eeq
where it is seen that the prefactor $\frac{1}{2!}$ exactly cancels with the two
in Eq. \ref{sum_con}. The contractions in Eq. \ref{2foldex} can therefore
be performed separately as shown in Eq. \ref{sum_con}. 
The exact same can be seen for a four-fold nested commutator for Eq. \ref{final_cc}
where, if the large basis set limit as shown in Eq. \ref{conord} is taken,
the contraction will be in the outer commutator, here $t_{a \alpha}^{(3)}$ 
and $t_{a \alpha}^{(4)}$, which can be rearranged to
\beq
\label{4foldex}
\hat C_{\alpha}^{dx1} \hat C_{\alpha}^{dx2} \hat t_{a \alpha}^4 =
\hat C_{\alpha}^{dx1} \hat C_{\alpha}^{dx2} \hat t_{a \alpha}^{(1)} \hat t_{a \alpha}^{(2)} \hat t_{a \alpha}^{(3)} \hat t_{a \alpha}^{(4)} =
\hat C_{\alpha}^{dx1} \hat C_{\alpha}^{dx2} \hat t_{a \alpha}^{(3)} \hat t_{a \alpha}^{(4)} \hat t_{a \alpha}^{(1)} \hat t_{a \alpha}^{(2)} (-1)^M 
\eeq
where again the contractions can be contracted separately as
shown in Eq. \ref{sum_con}. $\hat A_{\alpha}^{ex}$ is not
present in Eq. \ref{4foldex} since a four-fold nested
commutator must have four de-excitation terms and only
the Hamiltonian contain four de-excitation terms.

Since the Hamiltonian is index restricted the contractions
for unequal following contractions is
slightly different than for equal ones. The operator
$\hat H_{202}^{1,2}$ can be taken as an example for this.
For $\hat H_{202}^{1,2}$ the first contraction is 1,1
and the second 0,1 if the contraction order in Eq. \ref{conord}
is taken. If now the contraction in Eq. \ref{four_cc3} is performed
as in Eq. \ref{sum_con} many contraction would be missed
so here the contraction must be performed as shown in Eq. \ref{diff_con}.
The particle contraction from Eq. \ref{four_cc} will then 
have to be performed with $\hat t_{a\alpha}^{(1)}$ in this example
\beq
\contraction{}{\hat A}{_{\alpha}^{dx1} }{\hat t}
\hat A_{\alpha}^{dx1} \hat t_{a\alpha}^{(1)} \hat t_{a\alpha}^{(2)}
\eeq

The prefactor of $\frac{1}{n!}$ due to permutational symmetry 
can be therefore be completely removed by introducing
a contraction order and fixing the contractions as
shown in Eqs. \ref{sum_con} and \ref{diff_con} since this eliminates all
prefactors. This was also used in an earlier
implementation of a GASCC code\cite{krcc} although
in that code for practical reason the fixing of
the contraction order in Eq. \ref{sum_con} was
not used and therefore these contractions came
with a factor of $\frac{1}{2!}$.

\section{Algorithm}
%%%%%%%%%%%%%%%%%%%%%%%%%%%%%%%%%%%%%%%%%%%%%%%%
\label{SEC:imp}

Despite the three ways to perform the calculation
of the nested commutators presented in Sec. \ref{col_int2}
appear different the algorithm for these will 
all be very reminiscent of the ISCI algorithm
since the IS and IIS is obtained in the same way.
The similarities in performance between between collecting the
intermediates from a single integral and performing
all contraction in a single step when an IIS
is introduced is very small so only the latter
is shown.
The difference between the algorithms for collecting
intermediates and performing all contractions 
in a single step will here be shown and analyzed.

\subsection{Collecting intermediates}
\label{col_int_imp}

By collecting all intermediates an algorithm very similar to
that presented for the GASCC \cite{krcc} combined with
the ISCI can be constructed \cite{isci}. The contraction pattern
between the different $PHN\Delta M$-classes can then
be set up exactly like in the GASCC \cite{krcc} if the very
large basis set limit as shown in Eq. \ref{conord} is taken. Here
it is also possible to construct the intermediates piecewise
by using the GAS and in this way these can be made
almost arbitrarily small without any increase in the prefactor
or scaling. The number of times an integral will have to
be fetched or calculated will, however, increase.

As mentioned in Sec. \ref{riis} that for $n$ equal to zero
and one in Eqs. \ref{four_cc}-\ref{final_cc} the contractions are exactly like
for the ISCI with a slight modification. 
In fact the ISCI solution can be used
for all contraction directly pointing to $\Omega_{\hat T}$ since
in these contractions all indices are contracted.
The ISCI solution can, however, not be used directly when only
one of two de-excitation operators of the same kind
has to be contracted, as shown in Eqs. \ref{diff_con} and \ref{sum_con}, or when no 
contraction of a de-excitation operator
is performed. This happens for doubly and higher
nested commutators since not all indices in Eqs. \ref{four_cc}-\ref{final_cc}
are contracted in a single step like in the ISCI. 
The differences between the algorithm for 
these cases are minor as will be shown below.

For contractions pointing to $\Omega_{\hat T}$
the ISCI algorithm, where intermediate strings are constructed \cite{isci},
only needs very minor modification. Instead of 
contraction with the Hamiltonian, like in CI,
the contraction is performed with an operator
of general particle rank but with a maximum
down rank of two as described in Eq. \ref{iscc_ham}.
In the case where only one of two de-excitation
indices is contracted it is seen from Eq. \ref{sum_con}
it is sufficient to demand that only the last index $(dx2)$
is contracted if the following contraction is identical
while for non-identical Eq. \ref{diff_con} must 
be used. If no indices is contracted
a copy of the string of indices suffices.
\begin{algorithmic}[1]
\LOOP[ Strings $\hat A_{\alpha}^{dx}$ ]
 \LOOP[ Strings $\hat t_{c \alpha}$ ]
  \IF{ All indices contracted}
   \STATE{ Contract $\hat A_{\alpha}^{dx}$ with $\hat t_{c \alpha}$ to intermediate strings $\hat I$ with intermediate phase }
  \ELSIF{ One of two indices contracted}
   \IF{ The second contraction is identical }
    \STATE{ Contract $\hat A_{\alpha}^{dx2}$ with $\hat t_{c \alpha}$ to intermediate strings $\hat I$ with intermediate phase }
   \ELSIF{ The second contraction is not identical }
    \STATE{ Contract $\hat A_{\alpha}^{dx1}$ or $\hat A_{\alpha}^{dx2}$ with $\hat t_{c \alpha}$ to intermediate strings $\hat I$ with intermediate phase }
   \ENDIF{ Identical and non-identical contractions }
  \ELSIF{ No indices contracted}
   \STATE{ Copy $\hat t_{c \alpha}$ to intermediate strings $\hat I$ with intermediate phase }
  \ENDIF{ Number of indices contracted}
 \ENDLOOP[ Strings $\hat t_{c \alpha}$ ]
  \IF{ Any contraction between $\hat A_{\alpha}^{dx}$ and $\hat t_{c \alpha}$ is possible \ie number of $\hat I \ge 1$ }
   \LOOP[ Strings $\hat C_{\alpha}^{ex}$ ]
    \LOOP[ Strings $\hat I$ ]
     \STATE{ Add $\hat C_{\alpha}^{ex}$ to intermediate string $\hat I$ for final string $\hat O_{c \alpha}$ and phase }
     \IF{ Addition of $\hat C_{\alpha}^{ex}$ and $\hat I$ to $\hat O_{c \alpha}$ is possible }
      \STATE{ Calculate a relative offset for $\hat O_{c \alpha}$ string }
      \STATE{ Store relative offset from strings $\hat t_{c \alpha}$ and $\hat O_{c \alpha}$ }
      \STATE{ Store Hamiltonian indices from strings $\hat A_{\alpha}^{dx}$ and $\hat C_{\alpha}^{ex}$ }
      \STATE{ Store total phase for contraction and addition }
     \ENDIF{ Addition of $\hat C_{\alpha}^{ex}$ and $ \hat I$ to $\hat O_{c \alpha}$ is possible }
    \ENDLOOP[ Strings $\hat I$ ]
   \ENDLOOP[ Strings $\hat C_{\alpha}^{ex}$ ]
  \ENDIF{ Any contraction between $\hat A_{\alpha}^{dx}$ and $\hat t_{c \alpha}$ is possible }
\ENDLOOP[ Strings $\hat A_{\alpha}^{dx}$ ]
\end{algorithmic}
The loop structure is very similar to the ISCI 
where first the indices in a given $\hat A_{\alpha}^{dx}$
annihilation string are contracted with the ${\hat T}$ creation string
$\hat t_{c \alpha}$ to a set of intermediate creation strings $\hat I$.
Here a simple if-statement is inserted to separate if
all, one of two or no indices is contracted from the
de-excitation operator. Since the Hamiltonian and the intermediate
is index restricted there needs to be
a separation between identical and non-identical
following contractions as discussed in Sec. \ref{riis}. 
Identical following contractions
means that the same number of particle and hole indices
are contracted after each other as shown in Eq. \ref{H02_double}
while non-identical happens for contractions like
\beq
\hat H^{2,1} \otimes \hat T (11) \Rightarrow \hat M^{1,0} \otimes \hat T (10) \Rightarrow \Omega_{\hat T}
\eeq
where the particle index contraction is split in two.
After the contraction the creation strings $\hat C_{\alpha}^{ex}$ are then added to the intermediate
creation strings $\hat I$. The relative offsets from the
creation strings of $\hat t_{c \alpha}$
and $\hat O_{c \alpha}$, the operator 
indices in $\hat A_{\alpha}^{dx}$ and $\hat C_{\alpha}^{ex}$
along with a total phase for the contraction and addition of the strings are stored.
The operator $\hat O_{c \alpha}$ can here either be
an intermediate $\hat M_{c \alpha}$ or the projection manifold $\hat \Omega_{c \alpha}$
depending on where the contraction is pointing.
Just like for the ISCI is the first loop over the operator indices
crucial for a rigorous IIS. The major difference
in comparison to the ISCI is that not all
indices in $\hat A_{\alpha}^{dx}$ always
will be contracted.
For Eq. \ref{four_cc2} the
$\alpha$ spins in Eq. \ref{four_cc} are substituted with $\beta$ spins
and the same algorithm can then be used.
The loop structure for Eq. \ref{four_cc3} is:
\begin{algorithmic}[1]
\LOOP[ Strings $\hat A_{\alpha}^{ex}$ ]
 \LOOP[ Strings $\hat t_{a \alpha}$ ]
  \STATE{ Add $\hat A_{\alpha}^{ex}$ to $\hat t_{a \alpha}$ for intermediate strings $\hat I$ with intermediate phase }
 \ENDLOOP[ Strings $\hat t_{a \alpha}$ ]
 \IF{ Addition of $\hat A_{\alpha}^{ex}$ and $\hat t_{a \alpha}$ is possible \ie number of $\hat I \ge 1$}
  \LOOP[ Strings $\hat C_{\alpha}^{dx}$ ]
   \LOOP[ Strings $\hat I$ ]
   \IF{ All indices contracted}
    \STATE{ Contract $\hat C_{\alpha}^{dx}$ with $\hat I$ to final string $\hat O_{a \alpha}$ and phase }
   \ELSIF{ One of two indices contracted}
   \IF{ The second contraction is identical }
    \STATE{ Contract $\hat C_{\alpha}^{dx2}$ with $\hat I$ to final string $\hat O_{a \alpha}$ and phase }
   \ELSIF{ The second contraction is not identical }
    \STATE{ Contract $\hat C_{\alpha}^{dx1}$ or $\hat C_{\alpha}^{dx2}$ with $\hat I$ to final string $\hat O_{a \alpha}$ and phase }
   \ENDIF{ Identical and non-identical contractions }
   \ELSIF{ No indices contracted}
    \STATE{ Copy $\hat I$ to final string $\hat O_{a \alpha}$ and phase }
   \ENDIF{ Number of indices contracted}
   \IF{ Contraction of $\hat C_{\alpha}^{dx}$ and $\hat I$ to $\hat O_{a \alpha}$ is possible \ie number of $\hat O \ge 1$}
    \STATE{ Calculate a relative offset for $\hat O_{a \alpha}$ string }
    \STATE{ Store relative offset from strings $\hat t_{a \alpha}$ and $\hat O_{a \alpha}$ }
    \STATE{ Store Hamiltonian indices from string $\hat A_{\alpha}^{ex}$ and $\hat C_{\alpha}^{dx}$ }
    \STATE{ Store total phase for contraction and addition }
   \ENDIF{ Contraction of $\hat C_{\alpha}^{dx}$ and $\hat I$ to $\hat O_{a \alpha}$ is possible }
   \ENDLOOP[ Strings $\hat I$ ]
  \ENDLOOP[ Strings $\hat C_{\alpha}^{dx}$ ]
 \ENDIF{ Addition of $\hat A_{\alpha}^{ex}$ and $\hat t_{a \alpha}$ to intermediate string is possible }
\ENDLOOP[ Strings $\hat A_{\alpha}^{ex}$ ]
\end{algorithmic}
where the only difference to the loop structure for Eq. \ref{four_cc}
is the order in which the addition and contraction is performed.
For Eq. \ref{final_cc} we again can substitute $\alpha$ for $\beta$.
The strings in the contraction step are symbolically manipulated so the
creation and annihilation operator that should be contracted stand
next to each other, a sign for the number of transpositions is
calculated and the contracted indices are removed for the resulting $\hat O_{ax}$
string.

With the algorithms above Eqs. \ref{four_cc}-\ref{final_cc} can be 
solved for $n$ equal to zero or one which is all that is necessary
when intermediates are constructed. The aim here is to solve
the tensor contractions in a way similar to the ISCI
where there can be an IIS as shown in the algorithm below:
\begin{algorithmic}[1]
\LOOP[ Indices for $\hat O$ ]
 \STATE{Fetch or calculate integral $I$ and add to intermediate $M$}
 \IF{ $|I+M| > \epsilon $ }
 \LOOP[ Matrix elements ]
  \STATE{Multiply integral with element in $\hat v$ to element in $\hat O_f$}
 \ENDLOOP[ Matrix elements ]
 \ENDIF
\ENDLOOP[ Integral indices ]
\end{algorithmic}
Here the operator $\hat O_f$ is either a new intermediate $\hat M$
or the projection manifold $\hat \Omega_{\hat T}$.
Once the solution to Eqs. \ref{four_cc}-\ref{final_cc}
is known then a general loop structure where an integral
and or intermediate only will be fetched once and then
immediately multiplied with the amplitudes to a
generic operator $\hat O$
in a way very similar to the ISCI can be constructed.
\begin{algorithmic}[1]
\LOOP[ $\hat C_{\beta}^{dx} \hat A_{\beta}^{ex}$ ]
 \STATE{Get indices from $\hat C_{\beta}^{dx}$ and $\hat A_{\beta}^{ex}$ if needed}
 \STATE{Get number of $\hat t_{a \beta}$ strings and offset}
 \LOOP[ $\hat C_{\alpha}^{dx} \hat A_{\alpha}^{ex} $ ]
  \STATE{Get indices from $\hat C_{\alpha}^{dx} $ and $\hat A_{\alpha}^{ex} $ if needed}
  \STATE{Get number of $\hat t_{a \alpha}$ strings and offset}
  \LOOP[ $\hat C_{\beta}^{ex} \hat A_{\beta}^{dx} $ ]
   \STATE{Get indices from $\hat C_{\beta}^{ex} $ and $\hat A_{\beta}^{dx} $ if needed}
   \STATE{Get number of $\hat t_{c \beta}$ strings and offset}
   \LOOP[ $\hat C_{\alpha}^{ex} \hat A_{\alpha}^{dx} $ ]
    \STATE{Get indices from $\hat C_{\alpha}^{ex} $ and $\hat A_{\alpha}^{dx} $ if needed}
    \STATE{Get number of $\hat t_{c \alpha}$ strings and offset}
    \STATE{Fetch or calculate integral $I$ and add intermediate $M$}
    \IF{ $|I+M| > \epsilon $ }
    \LOOP[ $\hat t_{a \beta}$ ]
     \STATE{Get relative offsets and phase for $\hat O_{a \beta}$ and $\hat t_{a \beta}$}
     \LOOP[ $\hat t_{a \alpha}$ ]
      \STATE{Get relative offsets and phase for $\hat O_{a \alpha} $ and $\hat t_{a \alpha} $}
      \LOOP[ $\hat t_{c \beta}$ ]
       \STATE{Get relative offsets and phase for $\hat O_{c \beta} $ and $\hat t_{c \beta} $}
       \LOOP[ $\hat t_{c \alpha}$ ]
        \STATE{Get relative offsets and phase for $\hat O_{c \alpha} $ and $\hat t_{c \alpha} $}
        \STATE{Calculate total offset from relative offsets for $\hat O$ and $\hat T$}
        \STATE{Calculate total phase from relative phases and the overall phase}
        \STATE{Multiply integral with element in $\hat T$ and overall phase to element in $\hat O$}
       \ENDLOOP[ $\hat t_{c \alpha}$ ]
      \ENDLOOP[ $\hat t_{c \beta}$ ]
     \ENDLOOP[ $\hat t_{a \alpha}$ ]
    \ENDLOOP[ $\hat t_{a \beta}$ ]
    \ENDIF
   \ENDLOOP[ $\hat C_{\alpha}^{ex} \hat A_{\alpha}^{dx} $ ]
  \ENDLOOP[ $\hat C_{\beta}^{ex} \hat A_{\beta}^{dx} $ ]
 \ENDLOOP[ $\hat C_{\alpha}^{dx} \hat A_{\alpha}^{ex} $ ]
\ENDLOOP[ $\hat C_{\beta}^{dx} \hat A_{\beta}^{ex}$ ]
\end{algorithmic}
If the contraction is to the
projection manifold $\hat \Omega$ then the ISCI algorithm
can be used directly just by including the addition of the 
intermediate. If, however, the contraction is to an 
intermediate $\hat M$ the relative offset for $\hat O$ in the
inner loops also needs to be modified since the 
uncontracted de-excitation operator is included in the
outer loops. 

In the ISCI the general algorithm can be broken into 42 
different matrix-vector products. These products would
all also be present in the IISCC along with 61 additional
products, where an operator is contracted to an intermediate,
if the large basis set limit for the contractions in 
Eq. \ref{conord} is taken.

By constructing intermediates and introducing a contraction
order the IISCC will scale correctly and the prefactors can be 
minimized by the collection of intermediates as discussed in Sec. \ref{col_int}. 
The reduction in scaling
is therefore assumed to be very close to that seen for
the ISCI \cite{isci} once an IIS threshold is set. The
problem of storing the intermediates can be solved by
introducing many GAS since this reduces each block
that needs to be stored \cite{isci} at the expense
of a slightly more complicated algorithm. The main problem
for this approach occurs once more than doubles
is included since this increases the dimension
of the outer loops over the integrals and intermediates
which are the rate determining step in the IISCC for
large systems.
For very large systems it is, however, doubtful
that all full iterative triples or higher excitations can be
included due to the sheer number amplitudes.
Introducing a range dependent IIS will, just
like in the ISCI, reduce the size of the outer loop
but can also help to significantly reduce the
number of amplitudes and thereby making higher
than double excitations possible. 

%\subsection{Collecting intermediates from single integral} 

\subsection{All contractions in a single step} 

Taking the completely opposite approach where
no intermediates are collected and all contractions are
performed in a single step the IIS is still possible
for every multiplication in the nested commutators.
The algorithm will in many aspects be 
similar to the algorithm for collecting intermediates
in Sec. \ref{col_int_imp}. Here the contraction order
will in principle not matter but the arrangement of the loop
structure will. While the scaling of the nested commutators
is not correct this need not be a problem
since the IIS will be performed 
for the integrals and every time a
multiplication is performed in the nested
commutators. 

Exactly like in the algorithm which collects
intermediates presented in Sec. \ref{col_int_imp}
Eqs. \ref{four_cc}-\ref{final_cc} for $n$ equal
to zero or one can be solved using the ISCI algorithm.
For the nested commutators the ISCI algorithm
needs to be modified. Since the cluster operator
is a pure excitation operator it easy to reorder
this as shown in Eq. \ref{4foldex}. The contraction
can therefore be performed first and then 
added to the rest of the operators. To ensure that the
correct part of the Hamiltonian is contracted
with the cluster operator a contraction order
must be set. For the particle contractions in the algorithms below the large
basis set limit in Eq. \ref{conord} is taken.
\begin{algorithmic}[1]
\LOOP[ Strings $\hat A_{\alpha}^{dx}$ ]
 \STATE{ Reorder $\hat t_{c \alpha}^n$ so contractions are first with an overall phase, listed a,b,c,d } \label{all_l2}
 \IF{ Number of hole contractions is one }
  \LOOP[ Strings $\hat t_{c \alpha}^a$ ]
   \STATE{ Contract $\hat A_{\alpha}^{dx}$ with $\hat t_{c \alpha}^a$ to intermediate strings $\hat J$ with intermediate phase }
  \ENDLOOP[ Strings $\hat t_{c \alpha}^a$ ]
 \ELSIF{ Number of hole contractions is two}
  \IF{ The second contraction is identical }
   \LOOP[ Strings $\hat t_{c \alpha}^a$ ]
    \STATE{ Contract $\hat A_{\alpha}^{dx2}$ with $\hat t_{c \alpha}^a$ to intermediate strings $\hat I$ with intermediate phase }
    \LOOP[ Strings $\hat t_{c \alpha}^b$ ]
     \STATE{ Contract $\hat A_{\alpha}^{dx1}$ with $\hat t_{c \alpha}^b$ and add $\hat I$ to intermediate strings $\hat J$ with intermediate phase }
    \ENDLOOP[ Strings $\hat t_{c \alpha}^b$ ]
   \ENDLOOP[ Strings $\hat t_{c \alpha}^a$ ]
  \ELSIF{ The second contraction is not identical }
   \LOOP[ Strings $\hat t_{c \alpha}^a$ ]
    \STATE{ Contract $\hat A_{\alpha}^{dx1}$ or $\hat A_{\alpha}^{dx2}$ with $\hat t_{c \alpha}^a$ to intermediate strings $\hat I$ with intermediate phase }
    \LOOP[ Strings $\hat t_{c \alpha}^b$ ]
     \STATE{ Contract $\hat A_{\alpha}^{dx1}$ or $\hat A_{\alpha}^{dx2}$ with $\hat t_{c \alpha}^b$ and add $\hat I$ to intermediate strings $\hat J$ with intermediate phase }
    \ENDLOOP[ Strings $\hat t_{c \alpha}^b$ ]
   \ENDLOOP[ Strings $\hat t_{c \alpha}^a$ ]
  \ENDIF{ Identical and non-identical contractions }
 \ENDIF{ Number of indices contracted}
  \IF{ Any contraction with $\hat A_{\alpha}^{dx}$ is possible \ie number of strings $\hat J \ge 1$ }
   \LOOP[ Strings $\hat C_{\alpha}^{ex}$ ]
    \LOOP[ Strings $\hat J$ ]
     \STATE{ Add $\hat C_{\alpha}^{ex}$ to intermediate string $\hat J$ to intermediate string $\hat K$ } 
     \LOOP[ M-fold loop over remaining $\hat t_{c \alpha}^x$ ]
      \STATE{ Add $\hat t_{c \alpha}^x$ to intermediate string $\hat K$ for final string $\hat O_{c \alpha}$ and phase }
      \IF{ Addition of $\hat C_{\alpha}^{ex}$ and $\hat I$ to $\hat O_{c \alpha}$ is possible }
       \STATE{ Calculate a relative offset for $\hat O_{c \alpha}$ string }
       \STATE{ Store relative offset from strings $\hat t_{c \alpha}^n$ and $\hat O_{c \alpha}$ }
       \STATE{ Store Hamiltonian indices from strings $\hat A_{\alpha}^{dx}$ and $\hat C_{\alpha}^{ex}$ }
       \STATE{ Store total phase for contraction and addition }
      \ENDIF{ Addition of $\hat C_{\alpha}^{ex}$ and $ \hat I$ to $\hat O_{c \alpha}$ is possible }
     \ENDLOOP[ Remaining commutators ]
    \ENDLOOP[ Strings $\hat I$ ]
   \ENDLOOP[ Strings $\hat C_{\alpha}^{ex}$ ]
  \ENDIF{ Any contraction between $\hat A_{\alpha}^{dx}$ and $\hat t_{c \alpha}$ is possible }
\ENDLOOP[ Strings $\hat A_{\alpha}^{dx}$ ]
\end{algorithmic}
In order to perform the contractions first the $\hat t_{c \alpha}^n$
operators are reordered in line \ref{all_l2} to match the
contraction order in Eq. \ref{conord}. As shown in Eq. \ref{4foldex}
this only introduces an overall phase. Since the particles
are contracted before the holes in Eq. \ref{conord} the reorder
is only applicable when the de-excitation indices have different spins. After the contraction
of the de-excitation part, where all indices are contracted,
the excitation part is added and the remaining cluster operator
for which the contraction is in Eqs. \ref{four_cc2}-\ref{final_cc}.
Even though the algorithm above stores all possible contractions
the storage requirement for this can be dramatically reduced 
by introducing many GAS's \cite{lasse_unpub} exactly like in
the ISCI.

For the hole contractions the reorder in line \ref{all_h2} will be happen more
often since these contractions are usually performed
last.
\begin{algorithmic}[1]
\LOOP[ Strings $\hat A_{\alpha}^{ex}$ ]
 \STATE{ Reorder $\hat t_{a \alpha}^n$ so contractions are first with an overall phase, listed a,b,c,d } \label{all_h2}
 \IF{ Single contraction with $\hat t_{a \alpha}$} \label{all_h3}
  \LOOP[ Strings $\hat t_{a \alpha}^{(a)}$ ]
   \STATE{ Add $\hat A_{\alpha}^{ex}$ to $\hat t_{a \alpha}$ for intermediate strings $\hat I$ with intermediate phase }
  \ENDLOOP[ Strings $\hat t_{a \alpha}^{(a)}$ ]
 \ELSIF{ Double contraction with $\hat t_{a \alpha}$} \label{all_h7}
  \LOOP[ Strings $\hat t_{a \alpha}^{(a)}$ ]
   \STATE{ Add $\hat A_{\alpha}^{ex}$ to $\hat t_{a \alpha}^{(a)}$ for intermediate strings $\hat J$ with intermediate phase }
   \LOOP[ Strings $\hat t_{a \alpha}^{(b)}$ ]
    \STATE{ Add $\hat t_{a \alpha}^{(b)}$ to $\hat J$ for intermediate strings $\hat I$ with intermediate phase }
   \ENDLOOP[ Strings $\hat t_{a \alpha}^{(b)}$ ]
  \ENDLOOP[ Strings $\hat t_{a \alpha}^{(a)}$ ]
 \ENDIF
 \IF{ Addition of $\hat A_{\alpha}^{ex}$ and $\hat t_{a \alpha}$ is possible \ie number of $\hat I \ge 1$}                                               
  \LOOP[ Strings $\hat C_{\alpha}^{dx}$ ]                                                                                                                
   \LOOP[ Strings $\hat I$ ]                                                                                                                             
    \IF{ Single contraction with $\hat t_{a \alpha}$}                                
     \STATE{ Contract $\hat C_{\alpha}^{dx}$ with $\hat I$ to intermediate strings $\hat K$ and phase } 
    \ELSIF{ Double contraction with $\hat t_{a \alpha}$}                                                                                                                
     \IF{ The second contraction is identical }                                                                                                            
      \STATE{ Contract $\hat C_{\alpha}^{dx2}$ with $\hat t_{a \alpha}^{(a)}$ part in $\hat I$ to intermediate strings $\hat L$ and phase }
      \STATE{ Contract $\hat C_{\alpha}^{dx1}$ with $\hat t_{a \alpha}^{(b)}$ part in $\hat L$ to intermediate strings $\hat K$ and phase }
     \ELSIF{ The second contraction is not identical }                                                                                                     
      \STATE{ Contract $\hat C_{\alpha}^{dx1}$ or $\hat C_{\alpha}^{dx2}$ with with $\hat t_{a \alpha}^{(a)}$ part in $\hat I$ to intermediate string $\hat L$ and phase }                    
      \STATE{ Contract $\hat C_{\alpha}^{dx1}$ or $\hat C_{\alpha}^{dx2}$ with with $\hat t_{a \alpha}^{(b)}$ part in $\hat L$ to intermediate string $\hat K$ and phase }                    
     \ENDIF{ Identical and non-identical contractions }                                                                                                   
    \ENDIF{ Number of indices contracted}                                                                                                                 
    \LOOP[ M-fold loop over remaining $\hat t_{a \alpha}^x$ ] \label{downhere}
     \STATE{ Add $\hat t_{a \alpha}^x$ to intermediate string $\hat K$ for final string $\hat O_{a \alpha}$ and phase }
     \IF{ Contraction of $\hat C_{\alpha}^{dx}$ and $\hat I$ to $\hat O_{a \alpha}$ is possible \ie number of $\hat O_{a \alpha} \ge 1$}
      \STATE{ Calculate a relative offset for $\hat O_{a \alpha}$ string }                                                                                 
      \STATE{ Store relative offset from strings $\hat t_{a \alpha}^n$ and $\hat O_{a \alpha}$ }                                                             
      \STATE{ Store Hamiltonian indices from string $\hat A_{\alpha}^{ex}$ and $\hat C_{\alpha}^{dx}$ }                                                     
      \STATE{ Store total phase for contraction and addition }                                                                                             
     \ENDIF{ Contraction of $\hat C_{\alpha}^{dx}$ and $\hat I$ to $\hat O_{a \alpha}$ is possible }                                                       
    \ENDLOOP[ Remaining commutators ]
   \ENDLOOP[ Strings $\hat I$ ]                                                                                                                          
  \ENDLOOP[ Strings $\hat C_{\alpha}^{dx}$ ]                                                                                                             
 \ENDIF{ Addition of $\hat A_{\alpha}^{ex}$ and $\hat t_{a \alpha}$ to intermediate string is possible }                                                 
\ENDLOOP[ Strings $\hat A_{\alpha}^{ex}$ ]                                                                                                               
\end{algorithmic}                                                                                                                                        
For the hole contractions in lines \ref{all_h3} and \ref{all_h7}
an intermediate is built from either one or two cluster
operators which are later contracted by $\hat C_{\alpha}^{dx}$.
For the contraction it is again necessary to distinguish between
how following contractions are performed and the each de-excitation
term here is contracted with a specific part of the intermediate string.
Since the information of the original strings of $\hat t_{a \alpha}^{(a)}$
and $\hat t_{a \alpha}^{(b)}$ is readily available finding
the contraction in a part of the intermediate can be done just
by comparing $\hat C_{\alpha}^{dx}$ with $\hat t_{a \alpha}^{(a)}$
and $\hat t_{a \alpha}^{(b)}$. After the contraction the remaining
cluster operators $\hat t_{a \alpha}^x$ are looped over in line \ref{downhere}
in order to find the final operator $\hat O_{a \alpha}$ whereafter
the information from the additions and contractions are stored.

Once the solution to Eqs. \ref{four_cc}-\ref{final_cc} is tabulated
a general loop for up to four fold nested commutators can be
constructed:
\begin{algorithmic}[1]
\LOOP[ $\hat C_{\beta}^{dx} \hat A_{\beta}^{ex}$ ] \label{g1}
 \STATE{Get indices from $\hat C_{\beta}^{dx}$ and $\hat A_{\beta}^{ex}$ if needed}
 \STATE{Get number of $\hat t_{a \beta}^4$ strings and offset}
 \LOOP[ $\hat C_{\alpha}^{dx} \hat A_{\alpha}^{ex} $ ]
  \STATE{Get indices from $\hat C_{\alpha}^{dx} $ and $\hat A_{\alpha}^{ex} $ if needed}
  \STATE{Get number of $\hat t_{a \alpha}^4$ strings and offset}
  \LOOP[ $\hat C_{\beta}^{ex} \hat A_{\beta}^{dx} $ ]
   \STATE{Get indices from $\hat C_{\beta}^{ex} $ and $\hat A_{\beta}^{dx} $ if needed}
   \STATE{Get number of $\hat t_{c \beta}^4$ strings and offset}
   \LOOP[ $\hat C_{\alpha}^{ex} \hat A_{\alpha}^{dx} $ ] \label{g10}
    \STATE{Get indices from $\hat C_{\alpha}^{ex} $ and $\hat A_{\alpha}^{dx} $ if needed}
    \STATE{Get number of $\hat t_{c \alpha}^4$ strings and offset}
    \STATE{Fetch or calculate integral $I$}
    \IF{ $|I| > \epsilon_1 $ } \label{g14} 
    \LOOP[ $\hat t_{a \beta}^{(1)}$ ]
     \STATE{Get relative offsets for $\hat t_{a \beta}^{(1)}$}
     \LOOP[ $\hat t_{a \alpha}^{(1)}$ ]
      \STATE{Get relative offsets for $\hat t_{a \alpha}^{(1)} $}
      \LOOP[ $\hat t_{c \beta}^{(1)}$ ]
       \STATE{Get relative offsets for $\hat t_{c \beta}^{(1)} $}
       \LOOP[ $\hat t_{c \alpha}^{(1)}$ ]
        \STATE{Get relative offsets for $\hat t_{c \alpha}^{(1)} $}
        \STATE{Calculate total offset from relative offsets for $\hat T^{(1)}$}
        \STATE{Multiply integral with element in $\hat t^{(1)}$ to element $J$}
    \IF{ $|J| > \epsilon_2 $ } \label{g25}
    \LOOP[ $\hat t_{a \beta}^{(2)}$ ]
     \STATE{Get relative offsets for $\hat t_{a \beta}^{(2)}$}
     \LOOP[ $\hat t_{a \alpha}^{(2)}$ ]
      \STATE{Get relative offsets for $\hat t_{a \alpha}^{(2)} $}
      \LOOP[ $\hat t_{c \beta}^{(2)}$ ]
       \STATE{Get relative offsets for $\hat t_{c \beta}^{(2)} $}
       \LOOP[ $\hat t_{c \alpha}^{(2)}$ ]
        \STATE{Get relative offsets for $\hat t_{c \alpha}^{(2)} $}
        \STATE{Calculate total offset from relative offsets for $\hat T^{(2)}$}
        \STATE{Intermediate $J$ with element in $\hat t^{(2)}$ to element $K$}
    \IF{ $|K| > \epsilon_3 $ } \label{g36}
    \LOOP[ $\hat t_{a \beta}^{(3)}$ ]
     \STATE{Get relative offsets for $\hat t_{a \beta}^{(3)}$}
     \LOOP[ $\hat t_{a \alpha}^{(3)}$ ]
      \STATE{Get relative offsets for $\hat t_{a \alpha}^{(3)} $}
      \LOOP[ $\hat t_{c \beta}^{(3)}$ ]
       \STATE{Get relative offsets for $\hat t_{c \beta}^{(3)} $}
       \LOOP[ $\hat t_{c \alpha}^{(3)}$ ]
        \STATE{Get relative offsets for $\hat t_{c \alpha}^{(3)} $}
        \STATE{Calculate total offset from relative offsets for $\hat T^{(3)}$}
        \STATE{Intermediate $K$ with element in $\hat t^{(3)}$ to element $L$}
    \IF{ $|L| > \epsilon_4 $ } \label{g47}
    \LOOP[ $\hat t_{a \beta}^{(4)}$ ]
     \STATE{Get relative offsets and phase for $\hat O_{a \beta}$ and $\hat t_{a \beta}^{(4)}$}
     \LOOP[ $\hat t_{a \alpha}^{(4)}$ ]
      \STATE{Get relative offsets and phase for $\hat O_{a \alpha} $ and $\hat t_{a \alpha}^{(4)} $}
      \LOOP[ $\hat t_{c \beta}^{(4)}$ ]
       \STATE{Get relative offsets and phase for $\hat O_{c \beta} $ and $\hat t_{c \beta}^{(4)} $}
       \LOOP[ $\hat t_{c \alpha}^{(4)}$ ]
        \STATE{Get relative offsets and phase for $\hat O_{c \alpha} $ and $\hat t_{c \alpha}^{(4)} $}
        \STATE{Calculate total offset from relative offsets for $\hat O$ and $\hat T^{(4)}$}
        \STATE{Calculate total phase from relative phases and the overall phase}
        \STATE{Intermediate $L$ with element in $\hat t^{(4)}$ and overall phase to element in $\hat O$}
       \ENDLOOP[ $\hat t_{c \alpha}^{(4)}$ ]
      \ENDLOOP[ $\hat t_{c \beta}^{(4)}$ ]
     \ENDLOOP[ $\hat t_{a \alpha}^{(4)}$ ]
    \ENDLOOP[ $\hat t_{a \beta}^{(4)}$ ]
    \ENDIF{ Intermediate screening $L$ }
       \ENDLOOP[ $\hat t_{c \alpha}^{(3)}$ ]
      \ENDLOOP[ $\hat t_{c \beta}^{(3)}$ ]
     \ENDLOOP[ $\hat t_{a \alpha}^{(3)}$ ]
    \ENDLOOP[ $\hat t_{a \beta}^{(3)}$ ]
    \ENDIF{ Intermediate screening $K$ }
       \ENDLOOP[ $\hat t_{c \alpha}^{(2)}$ ]
      \ENDLOOP[ $\hat t_{c \beta}^{(2)}$ ]
     \ENDLOOP[ $\hat t_{a \alpha}^{(2)}$ ]
    \ENDLOOP[ $\hat t_{a \beta}^{(2)}$ ]
    \ENDIF{ Intermediate screening $J$ }
       \ENDLOOP[ $\hat t_{c \alpha}^{(1)}$ ]
      \ENDLOOP[ $\hat t_{c \beta}^{(1)}$ ]
     \ENDLOOP[ $\hat t_{a \alpha}^{(1)}$ ]
    \ENDLOOP[ $\hat t_{a \beta}^{(1)}$ ]
    \ENDIF{ Integral screening $I$ }
   \ENDLOOP[ $\hat C_{\alpha}^{ex} \hat A_{\alpha}^{dx} $ ]
  \ENDLOOP[ $\hat C_{\beta}^{ex} \hat A_{\beta}^{dx} $ ]
 \ENDLOOP[ $\hat C_{\alpha}^{dx} \hat A_{\alpha}^{ex} $ ]
\ENDLOOP[ $\hat C_{\beta}^{dx} \hat A_{\beta}^{ex}$ ]
\end{algorithmic}
Exactly like for the ISCI and when collecting intermediates
in the IISCC the outer loops from lines \ref{g1}-\ref{g10} 
are over the indices in the Hamiltonian. An integral is
fetched or calculated and screened in line \ref{g14}.
After the IS is a series of loops over the nested
commutators with one cluster operator at a time follows. After
the first loops over a cluster operator the integral is
multiplied with an amplitude and screened as shown in line \ref{g25}.
A similar multiplication is performed after every
cluster operator and
in this way is not only the integrals screened but
also the intermediates. While the loop structure in the general algorithm
gives the wrong scaling the IIS in lines \ref{g14}, \ref{g25}, \ref{g36} and \ref{g47}
will reduce the scaling for spatially extended systems.
Since the numerical value of the amplitudes should
be below one it is possible to obtain the same
accuracy even while using a declining screening threshold
so the outer IIS threshold is higher than the inner IIS
\beq
|\epsilon_1| \ge |\epsilon_2| \ge |\epsilon_3| \ge |\epsilon_4| . 
\eeq
A declining IIS threshold will give a further speed up without destroying the
numerical accuracy.

From the general loop structure for a four-fold nested
commutator it is easy to construct three- and two-fold
nested commutators. For the remaining terms the algorithm
is exactly like the ISCI algorithm.
Since the IIS should reduce the scaling of the nested commutators
it is expected that these will not significantly increase
the scaling of the IISCC and that the performance will
be similar to that shown for the ISCI \cite{isci}. The
advantage of not collecting the intermediates is that
the outer loop will not depend on the excitation level
included in the CC expansion which will be beneficial
when triples or higher excitations are included.

%\section{Application}
%%%%%%%%%%%%%%%%%%%%%%%%%%%%%%%%%%%%%%%%%%%%%%%%
%\label{SEC:appl}
%\input{app}

\section{Summary and prospects}
%%%%%%%%%%%%%%%%%%%%%%%%%%%%%%%%%%%%%%%%%%%%%%%%
\label{SEC:summ}

We have here presented the derivation of the 
integral- and intermediate-screened coupled-cluster method (IISCC)
in which an \map combined integral and intermediate screening (IIS)
of the integrals and intermediates is expected to
significantly reduces the scaling in comparison to the regular CC method
as seen in the integral-screened configuration-interaction method (ISCI) \cite{isci}.
The simple and rigorous IIS ensures a good error
control and will allow for the
convergence of the energy to a very high accuracy
while still retaining a very low scaling. The
IIS only relies on numerical screening and 
does not contain any distance dependent screening
so the IISCC is equally suited for both neutral and charged
systems since both
the short and long range of the wave function is equally
well described. Due to the great similarities with the
ISCI it is expected that the IISCC will exhibit
the same low scaling properties for spatially extended
systems but unlike the ISCI be size extensive and therefore
also useful for many electron systems.

Currently the Hamiltonian is written in terms of index restricted and                                                                                            
normal- and spin-ordered operators 
since this was shown to give
a simple and rigorous IS in the ISCI.
Following the derivation of the ISCI and expressing the Hamiltonian in  
strings of second quantized operators all contractions of                         
all orders of nested commutators are separable up to                                                                                           
a sign. This separability allows for a 
general loop structure where in the
tensor contractions
the outer loops are over the indices                                                                                                               
of the integrals in the Hamiltonian. Having                        
the outer loops over the integrals allows                                                                                                                
for a very simple, efficient and rigorous \map IIS                                                                                                        
where only integrals and intermediates above a predefined threshold                                                                                                        
are computed.                                                                                                                  
As shown for the ISCI such a procedure will automatically give             
a reduction in the scaling for                                                                                                        
spatially extended systems in local basis sets
where linear scaling in the multiplication of
integrals and coefficients step is gradually
approached.

Three different approaches, where the algorithm
is shown for two of them, which all 
have a simple and rigorous IIS is presented.
In the first example, where intermediates are
collected exactly like in the standard CC method,
a simple and rigorous IIS can be obtained
simply by first looping over the combined
indices for the Hamiltonian and the intermediate.
The algorithm in the very large basis set limit
is therefore very close to a combination of
the algorithm presented for the generalized-active space
coupled-cluster method (GASCC) \cite{krcc} and the ISCI \cite{isci}.
The formal scaling and the prefactors will be
like that of the regular CC method but these
are expected to be reduced significantly
by the IIS like seen for the ISCI. While the size of the intermediates
grows significantly with system size and excitation level
a piecewise construction of these can reasonably
easy be obtained by introducing many GAS's
and storing these should therefore not be a problem.
In the ISCI the most expensive step quickly
becomes the loop over the integrals in the Hamiltonian
and this is also expected to be the most expensive
step for the IISCC. When only doubles is included
then the addition of the intermediates to the integrals will
not increase the cost of the outer loops, however,
once triples or higher is included then this loop
will increase significantly since intermediates
with a particle rank of three will appear.
Since the nested contractions are separable
up to a sign a completely alternative
route where all contractions in the nested commutators
are contracted in a single step is presented. In this
the intermediates are not explicitly constructed
and the outer loops will therefore not
be more expensive when higher than doubles excitations
are included. The scaling of this approach
for the nested commutators is higher than 
when intermediates is collected. The higher
scaling is not expected to pose a problem
since a gradual IIS in the nested commutators
can be used which means even more integrals
can in the outer step be safely discarded
without any loss of accuracy. The scaling
of the nested commutators will therefore
decrease even faster than the direct
contractions.
For both approaches
much of the machinery developed for the ISCI appear                                                                                                                  
to be directly applicable to the IISCC method. 
If the intermediates are collected an
additional 61 tensor contractions will
have to be implemented while for the
single step contraction the double, triple
and quadruple nested commutators should
be implemented. When the CC hierarchy
is truncated at the doubles level it
will clearly be advantages to collect
the intermediates while if more than doubles
is included it will be advantages to perform
all contractions in a single step as long
as no distant dependent truncation of the 
interaction is introduced.

While the main aim of the ISCI method was the accurate simulations of                                                                                          
physical processes where one or more electrons move in the continuum for                                                                                 
atoms and molecules in strong laser fields the aim of the IISCC extends beyond this
due to the size-extensive nature of the CC equations.
The IISCC would therefore also be very interesting with in electronic
structure theory
for large system with many electrons either
as a stand alone low scaling method or
in combination with the fragment based linear scaling
approach or the distance screening approach.
The central advantage of the IISCC is the ability
to use extremely large basis sets, as shown for
the ISCI, which is a central problem in the 
current methods used to obtain linear scaling.
Due to the rigorous IIS and error control 
the IISCC is expected to significantly 
reduce the errors for linearly scaling methods.

%%%%%%%%%%%%%%%%%%%%%%%%%%%%%%%%%%%%%%%%%%%%%%%%
%\input{ack}

%%%%%%%%%%%%%%%%%%%%%%%%%%%%%%%%%%%%%%%%%%%%%
%                                           %
%         Appendices                        %
%                                           %
%%%%%%%%%%%%%%%%%%%%%%%%%%%%%%%%%%%%%%%%%%%%%
\begin{appendix}
%%%%%%%%%%%%%%%%%%%%%%%%%%%%%%%%%%%%%%%%%%%%%%%%
\section{Excitation-class Formalism}
\label{ex_class_form}

In this section we recapitulate the excitation class formalism
for a non-relativistic Hamiltonian, as used in the ISCI \cite{isci},
and originally presented in the relativistic framework
for GASCC implementations \cite{soerensen_commcc,krcc}.
The excitation class formalism
maps a normal-ordered operator, consisting of a string of second
quantized operators, onto a set of classes helpful
in characterizing different parts of an operator and
in aiding the algorithm due to their simple algebra.
We will assume that the orbitals
have been optimized in some restricted way so that the
orbitals can be related by the spin-flip operator \cite{krcc}.

The $PHN\Delta M$-classes introduces a set of
auxiliary quantum numbers which depends only on
the number of alpha and beta creation and annihilation operators,
$N^c_{\alpha}, N^c_{\beta},  N^a_{\alpha}$ and $N^a_{\beta}$, respectively,
and hence have four indices where each preceding index
represent a further division of the classes.
Since we here are concerned with number conserving operators
\beq
\label{ncons}
N^{c \alpha} + N^{c \beta} = N^{a \alpha} + N^{a \beta},
\eeq
this leaves three additional indices which are chosen
as the particle rank $N$, spin flip $\Delta M_s$ and
the difference between the number of $\alpha$ and $\beta$
operators $M_{\alpha \beta}$.
The operator classes from a general operator like the Hamiltonian $\hat H$,
the excitation operator $\hat X$, intermediate $\hat M$ or any other number-conserving 
normal-ordered second quantized operators
can all be divided in the $PHN\Delta M$-classes in the same way
\beq
\label{mixop2}
\hat O = \sum_{P,H} \sum_{N,\Delta ,M}  \hat O_{N,\Delta , M}^{P,H}.
\eeq
Here $P$ and $H$ denote the de-excitation part,
$P$ gives the number of annihilation de-excitation terms and
$H$ the number of creation de-excitation terms
$N$ is the particle rank 
\beqa
\label{particle_rank}
N &=& \frac{1}{2}( N^{c \alpha} + N^{c \beta} + N^{a \alpha} + N^{a \beta} ) \nonumber \\
  &=& N^{c \alpha} + N^{c \beta} \quad \Rightarrow N^{c \alpha} + N^{c \beta} = N^{a \alpha} + N^{a \beta},
\eeqa
$\Delta$ is the spin flip of the spin orbitals \cite{jensen_saue,krcc} 
\beqa
\label{krflip}
\Delta M_s &=& \frac{1}{2}( N^{c \alpha} - N^{c \beta} + N^{a \alpha} - N^{a \beta} ) \nonumber \\                                  
           &=& N^{c \alpha} - N^{a \alpha} \quad \Rightarrow N^{c \alpha} + N^{c \beta} = N^{a \alpha} + N^{a \beta},
\eeqa
and $M_{\alpha \beta}$ is the difference in the number of operators with alpha and beta spins
\beqa
M_{\alpha \beta} &=& \frac{1}{2}( N^{c \alpha} - N^{c \beta} + N^{a \alpha} - N^{a \beta} ) \nonumber \\                            
                 &=& N^{c \alpha} - N^{a \beta} \quad \Rightarrow N^{c \alpha} + N^{c \beta} = N^{a \alpha} + N^{a \beta}.
\eeqa
While the Hamiltonian and the intermediate will have classes with non-zero $P$ and $H$
the excitation operator $\hat X$ will not, since $\hat X$ does not contain any
de-excitation terms according to the definition of
excitation and de-excitation operators in Section \ref{SEC:theo}.

\end{appendix}

%%%%%%%%%%%%%%%%%%%%%%%%%%%%%%%%%%%%%%%%%%%%%
%                                           %
%         Bibliography                      %
%                                           %
%%%%%%%%%%%%%%%%%%%%%%%%%%%%%%%%%%%%%%%%%%%%%
%\bibliographystyle{unsrt}
\bibliography{full}
%\bibliography{all}

\end{document}